\documentclass
[aps,prl,amsfonts,amssymb,twocolumn,amsmath,preprintnumbers,nofootinbib,floatfix,
showpacs]{revtex4-1}%
\usepackage[dvips]{graphics}
\usepackage{graphicx}
\usepackage{bm}
\usepackage{amsmath}
\usepackage{amsfonts}
\usepackage{amssymb}
\usepackage{xcolor}
\usepackage{subfigure}
\usepackage{hyperref,hypcap}
\usepackage{braket}
\usepackage{commath}%
\providecommand{\U}[1]{\protect\rule{.1in}{.1in}}

\begin{document}

\title{Berry-phase effects in dipole density and Mott relation}
\author{Liang Dong}
\thanks{These authors contributed equally to this work.}
\author{Cong Xiao}
\thanks{These authors contributed equally to this work.}
\email{Corresponding author: congxiao@utexas.edu}
\author{Bangguo Xiong}
\author{Qian Niu}
\affiliation{Department of Physics, The University of Texas at Austin, Austin, Texas 78712, USA}

\begin{abstract}
We provide a unified semiclassical theory for thermoelectric responses of any
observable represented by an operator $\hat{\boldsymbol{\theta}}$ that is
well-defined in periodic crystals. The Einstein and Mott relations are
established generally, in the presence of Berry-phase effects, for various
physical realizations of $\hat{\boldsymbol{\theta}}$ in electronic systems,
including the familiar case of the electric current as well as the currently
controversial cases of the spin polarization and spin current. The
magnetization current, which has been proven indispensable in the
thermoelectric response of electric current, is generalized to the cases of
various $\hat{\boldsymbol{\theta}}$. In our theory the dipole density of a
physical quantity emerges and plays a vital role, which contains not only the
statistical sum of the dipole moment of $\hat{\boldsymbol{\theta}}$ but also a
Berry-phase correction.
\end{abstract}
\maketitle


The Mott relation \cite{Ziman1960,Streda1977} was originally proposed as a
fundamental link between the measurable electric current responses to the
electric field and to the temperature gradient in independent-electron systems
with elastic scattering off static disorder. Since the rapid extension of the
fields of spintronics and spin-caloritronics \cite{Bauer2012}, the question
whether the Mott relation still holds for thermoelectric responses related to
the electronic spin degree of freedom in spin-orbit coupled systems has
attracted intensive debates \cite{Xiao2006,Miyasato2007,Pu2008}. In
particular, despite the recent experimental observation of the spin Nernst
effect \cite{Meyer2017,Sheng2017,Kim2017,Bose2018} the thermal
counterpart of the spin Hall effect \cite{Sinova2015,Nagaosa2008}, the puzzle
whether the Mott relation exists between these two effects has not been
settled theoretically
\cite{Ma2010,Dyrdal2016,Borge2013,SNE2014,Tauber2012,Xiao2018}. Besides,
whether the Edelstein effects (nonequilibrium spin polarization) induced by
the electric field \cite{Edelstein1990,Sinova2015} and by the
temperature-gradient \cite{Wang2010} are linked by the Mott relation is also a
controversial issue \cite{Dyrdal2018,Shitade2018}.

In the presence of band-structure spin-orbit coupling, various Berry-phase
effects on thermoelectric responses appear
\cite{Xiao2010,Nagaosa2010,Sinova2015,Nagaosa2008,Freimuth2014}. In
particular, the identification of the orbital magnetization including a
Berry-phase correction has been proven vital in validating the Mott relation
between the anomalous Nernst and anomalous Hall effects in ferromagnets
\cite{Xiao2006}. In this Letter we provide a unified semiclassical theory for
thermoelectric responses of any observable represented by an operator
$\hat{\boldsymbol{\theta}}$ that is well-defined in periodic crystals. We
establish the Einstein and Mott relations in the presence of Berry-phase
effects for various physical realizations of $\hat{\boldsymbol{\theta}}$,
including the known case of the electric current \cite{Xiao2006}, as well as
the intensively debated cases of the conventional spin current (defined as the
anti-commutator of the spin and velocity operators)
\cite{Ma2010,Dyrdal2016,Borge2013,SNE2014,Tauber2012,Xiao2018} and the spin
polarization \cite{Dyrdal2018,Shitade2018}. The magnetization current, which
has been proven indispensable in the thermoelectric response of electric
current \cite{Xiao2006}, is generalized to various $\hat{\boldsymbol{\theta}}%
$. As a generalization of the orbital magnetization in the case of the
electric current, in our theory the dipole density of a physical quantity
($\hat{\boldsymbol{\theta}}$) emerges and plays a vital role. It contains not
only the statistical sum of the dipole moment of $\hat{\boldsymbol{\theta}}$
\cite{Culcer2004} but also a Berry-phase correction.

In the strategy of the semiclassical theory \cite{Ashcroft}, one considers a
grand canonical ensemble of dynamically independent semiclassical Bloch
electrons, each of which is physically identified as a wave-packet
$|\Phi\left(  \boldsymbol{q}_{c},\boldsymbol{r}_{c},t\right)  \rangle$ that is
constructed from the Bloch states in a particular nondegenerate band and is
localized around a central position $\boldsymbol{r}_{c}$ and a mean crystal
momentum $\boldsymbol{q}_{c}$. Within the validity of the uncertainty
principle, the phase-space occupation function $f_{tot}\left(  \boldsymbol{q}%
_{c},\boldsymbol{r}_{c},t\right)  $ can be defined, and the density-of-states
$D\left(  \boldsymbol{q}_{c},\boldsymbol{r}_{c}\right)  $ has to be introduced
\cite{Xiao2005}. The number of states within a small phase space volume is
hence given by $Df_{tot}d\boldsymbol{r}_{c}d\boldsymbol{q}_{c}/(2\pi)^{3}$.
$f_{tot}=f+\delta f$, where $f$ is the local equilibrium Fermi distribution,
and $\delta f$ is a small deviation originating from scattering processes.

In this paper, we consider Bloch electrons in a crystal under small electric
field, spatially inhomogeneous chemical potential and temperature. We keep our
result to the first order of the gradients of the electrostatic potential and
chemical potential $\mu$ as well as temperature $T$. The electron wave-packet
in such a system is described by the following Hamiltonian:%
\begin{equation}
\hat{H}=\hat{H}_{0}\left(  \hat{\boldsymbol{p}}+\boldsymbol{q}_{c}%
,\hat{\boldsymbol{r}}\mathbf{;}w\left(  \boldsymbol{r}_{c}\right)  \right)
-e\phi\left(  \boldsymbol{r}_{c}\right)  , \label{eq: original Hamiltonian}%
\end{equation}
where the electrostatic potential $\phi\left(  \boldsymbol{r}_{c}\right)
=-\boldsymbol{E}\cdot\boldsymbol{r}_{c}$ is explicitly shown with
$\boldsymbol{E}$ the electric field, and $w\left(  \boldsymbol{r}_{c}\right)
$ represent other possible mechanical perturbation fields \cite{Sundaram1999}.
We focus on the static case such that $w\left(  \boldsymbol{r}_{c}\right)  $
does not depend on time. All these fields vary slowly on the scale of the
wave-packet. Thus their original $\hat{\boldsymbol{r}}$ dependence is replaced
by the $\boldsymbol{r}_{c}$ dependence under the local approximation. The
eigenstate of $\hat{H}$ is the same as that of $\hat{H}_{0}$ while the
eigenenergy is shifted by $-e\phi\left(  \boldsymbol{r}_{c}\right)  $. We
denote $\varepsilon(\boldsymbol{q}_{c},\boldsymbol{r}_{c})$ and
$|u(\boldsymbol{q}_{c},\boldsymbol{r}_{c})\rangle$ as the eigenenergy and
eigenstate (periodic part of Bloch function) of $\hat{H}_{0}$. Then the
phase-space density-of-states reads: $D\left(  \boldsymbol{q}_{c}%
,\boldsymbol{r}_{c}\right)  =1+\Omega_{q_{ci}r_{c}^{i}}$ \cite{Xiao2005},
where $\Omega_{\lambda_{i}\lambda_{j}}=2\operatorname{Im}\langle
\partial_{\lambda_{j}}u|\partial_{\lambda_{i}}u\rangle$ are the Berry
curvatures, $\lambda_{i}=r_{c}^{i}$ or $q_{c,i}$, $i$ and $j$ are Cartesian
indices. Summation over repeated indices is implied henceforth.

The local density of a physical observable $\hat{\boldsymbol{\theta}}$
(generally a tensor operator) is defined as \cite{Culcer2004}%
\begin{equation}
\rho_{tot}^{\boldsymbol{\theta}}\left(  \boldsymbol{r}\right)  \equiv
\int\left[  d\boldsymbol{q}_{c}\right]  d\boldsymbol{r}_{c}Df_{tot}\langle
\Phi|\hat{\boldsymbol{\theta}}\delta\left(  \hat{\boldsymbol{r}}%
-\boldsymbol{r}\right)  |\Phi\rangle. \label{eq: local density original}%
\end{equation}
We further divide it into two parts: $\rho_{tot}^{\boldsymbol{\theta}}%
=\rho_{loc}^{\boldsymbol{\theta}}+\delta\rho_{loc}^{\boldsymbol{\theta}}$,
where $\rho_{loc}^{\boldsymbol{\theta}}$ is the contribution from $f$ and
$\delta\rho_{loc}^{\boldsymbol{\theta}}$ is from $\delta f$. In the following,
we focus on $\rho_{loc}^{\boldsymbol{\theta}}$ while the discussion of
$\delta\rho_{loc}^{\boldsymbol{\theta}}$ is postponed to the end of the paper.
Hereafter the symmetrization between operators that do not commutate to each
other is implied. $\left[  d\boldsymbol{q}_{c}\right]  $ is shorthand for
$\sum_{n}d\boldsymbol{q}_{c}/\left(  2\pi\right)  ^{d}$ with $d$ the spatial
dimensionality (we use the convention $\hbar=1$). First-order Taylor expansion
of $\hat{\boldsymbol{r}}$ with respect to $\boldsymbol{r}_{c}$ in the Dirac
delta function yields \cite{Xiao2010}%
\begin{align}
\rho_{loc}^{\boldsymbol{\theta}}\left(  \boldsymbol{r}\right)  =  &
\int\left[  d\boldsymbol{q}_{c}\right]  Df\langle\Phi|\hat{\boldsymbol{\theta
}}|\Phi\rangle|_{\boldsymbol{r}_{c}=\boldsymbol{r}}\nonumber\\
&  -\boldsymbol{\nabla}\cdot\int\left[  d\boldsymbol{q}_{c}\right]
f\langle\Phi|\hat{\boldsymbol{\theta}}\left(  \hat{\boldsymbol{r}%
}-\boldsymbol{r}\right)  |\Phi\rangle|_{\boldsymbol{r}_{c}=\boldsymbol{r}},
\label{eq: local density}%
\end{align}
which is the basis of the following discussion. Henceforth we will omit the
center position label $c$, and the notation $\int$ without integral variable
is shorthand for $\int\left[  d\boldsymbol{q}_{c}\right]  $, unless otherwise
noted. We consider $\rho_{loc}^{\boldsymbol{\theta}}\left(  \boldsymbol{r}%
\right)  $ up to the first order, thus it is sufficient to set $D=1$ in the
second term of $\rho_{loc}^{\boldsymbol{\theta}}\left(  \boldsymbol{r}\right)
$. This term is related to the dipole moment of $\hat{\boldsymbol{\theta}}$
\cite{Culcer2004,Xiao2010}:%
\begin{equation}
m^{i\boldsymbol{\theta}}=\langle\Phi|\hat{\boldsymbol{\text{$\theta$}}}%
(\hat{\boldsymbol{r}}-\boldsymbol{r})^{i}|\Phi\rangle,
\end{equation}
whose physical meaning is shown in Fig.1. Whereas the first term of
$\rho_{loc}^{\boldsymbol{\theta}}\left(  \boldsymbol{r}\right)  $ is just the
conventional semiclassical expression \cite{Ziman1960}. \begin{figure}[tbh]
\includegraphics[width=0.7\columnwidth]{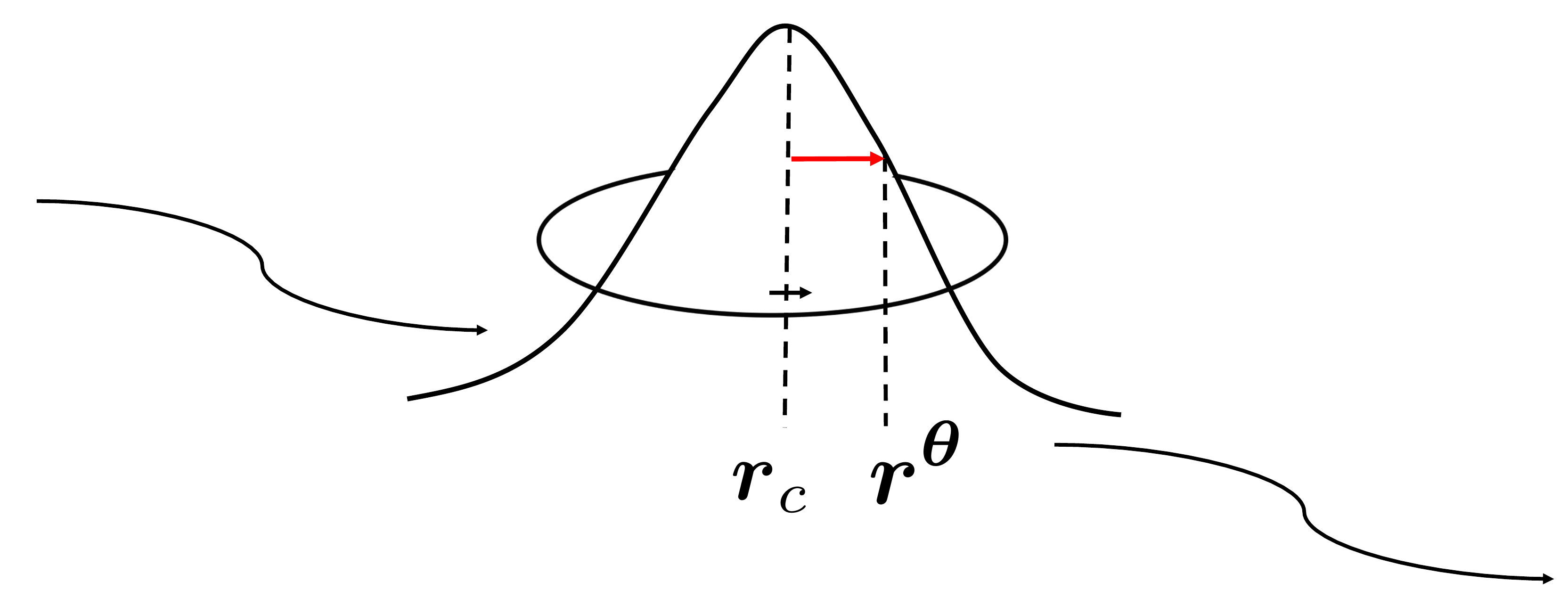}\caption{A schematic
picture of $m^{i\boldsymbol{\theta}}$ which is proportional to the difference
between the $\boldsymbol{\theta}$-center $\boldsymbol{r}^{\boldsymbol{\theta}%
}$ and the usual probability center $\boldsymbol{r}_{c}$ (the red arrow),
where the $\boldsymbol{\theta}$-center is defined as $\boldsymbol{r}%
^{\boldsymbol{\theta}}\equiv\frac{\langle\Phi|\hat{\boldsymbol{r}}%
\hat{\boldsymbol{\theta}}|\Phi\rangle}{\langle\Phi|\hat{\boldsymbol{\theta}%
}|\Phi\rangle}$. By definition (\ref{eq: local density}),
$m^{i\boldsymbol{\theta}}=\langle\Phi|\hat{\boldsymbol{\theta}}|\Phi
\rangle(\boldsymbol{r}^{\boldsymbol{\theta}}-\boldsymbol{r}_{c})^{i}$. For
\textquotedblleft conserved\textquotedblright\ $\hat{\boldsymbol{\theta}}$
that commutes with the Hamiltonian $\hat{H}_{0}$, the $\boldsymbol{\theta}%
$-center coincides with the probability center such that the
$\boldsymbol{\theta}$ dipole moment vanishes.}%
\label{figure}%
\end{figure}

Given the complexity of the present subject, we first look into the special
case when $\hat{\boldsymbol{\theta}}=-e\hat{\boldsymbol{v}}$ and the electric
current is calculated. Here we stress that, as will be discussed later, the
following derivation is not a repetition of what has been done in Ref.
\cite{Xiao2006}, but is a novel approach and provides a different perspective
which eventually inspires a general method applicable to observables other
than the electric current. The case of electric current is special because,
the first term in Eq. (\ref{eq: local density}) is now simply $-e\langle
\Phi|\hat{\boldsymbol{v}}|\Phi\rangle=-e\dot{\boldsymbol{r}}$ where
$\dot{\boldsymbol{r}}$ is the velocity of the wave-packet and is given by the
equations of motion \cite{Sundaram1999}
\begin{align}
\dot{\boldsymbol{r}}=  &  \partial_{\boldsymbol{q}}\varepsilon_{tot}%
-\Omega_{\boldsymbol{q}\mathcal{T}},\nonumber\\
\dot{\boldsymbol{q}}=  &  -\partial_{\boldsymbol{r}}\varepsilon_{tot}%
-e\boldsymbol{E}+\Omega_{\boldsymbol{r}\mathcal{T}}.
\label{eq: equations of motion}%
\end{align}
$\varepsilon_{tot}$ is the total wave-packet energy from $\hat{H}_{0}$:
$\varepsilon_{tot}=\varepsilon+\Delta\varepsilon$ with $\Delta\varepsilon
=\operatorname{Im}\langle\partial_{q_{ci}}u|\varepsilon-\hat{H}|\partial
_{w}u\rangle\partial_{r_{c}^{i}}w$ the contribution from the gradient of
$w\left(  \boldsymbol{r}_{c}\right)  $ \cite{Sundaram1999}. The Berry
curvature term reads $\Omega_{\lambda\mathcal{T}}=\Omega_{\lambda
\boldsymbol{r}}\dot{\boldsymbol{r}}+\Omega_{\lambda\boldsymbol{q}}%
\dot{\boldsymbol{q}}+\Omega_{\lambda t}$ with $\mathcal{T}$ denoting the total
time ($t$) derivative. $\Omega_{\lambda t}$ vanishes in the static case
studied here.

Therefore, the local electric current density $\boldsymbol{j}_{loc}$ reads%
\begin{equation}
\boldsymbol{j}_{loc}=-e\int Df(\partial_{\boldsymbol{q}}\varepsilon
_{tot}-\Omega_{\boldsymbol{q}\mathcal{T}})+\boldsymbol{\nabla}\times\int
f\boldsymbol{m}. \label{eq: local j_e}%
\end{equation}
Here $\boldsymbol{m}=e\operatorname{Im}\langle\partial_{\boldsymbol{q}%
}u|\times(\varepsilon-\hat{H})|\partial_{\boldsymbol{q}}u\rangle$ is the
vector form of the antisymmetric tensor $m^{ij}$ with the index $j$ coming
from the three components of $\hat{\boldsymbol{\theta}}=-e\hat{\boldsymbol{v}%
}$, and is known as the orbital magnetic moment of the wave-packet
\cite{Sundaram1999}. Substituting the equations of motion into the
$\Omega_{\boldsymbol{q}\mathcal{T}}$ term, after some algebra \cite{supp} we
find that $\boldsymbol{j}_{loc}$ can be divided into two parts:
\begin{equation}
\boldsymbol{j}_{loc}=\boldsymbol{j}_{eq}+\boldsymbol{j}_{neq}. \label{j}%
\end{equation}
The equilibrium part $\boldsymbol{j}_{eq}$ exists irrespective of the electric
field and statistical force (temperature gradient and chemical potential
gradient), while the non-equilibrium part $\boldsymbol{j}_{neq}$ is induced by
them. The two parts read:%
\begin{equation}
\boldsymbol{j}_{eq}=\boldsymbol{\nabla}\times\boldsymbol{M},\text{
\ \ }\boldsymbol{j}_{neq}=\boldsymbol{\sigma}^{i}\left(  E_{i}+\partial_{i}%
\mu/e\right)  -\boldsymbol{\alpha}^{i}\partial_{i}T,
\end{equation}
where the Hall and Nernst conductivities are given by $\boldsymbol{\sigma}%
^{i}=e^{2}\int f\Omega_{q_{i}\boldsymbol{q}}$ and $\boldsymbol{\alpha}%
^{i}=-\frac{e}{T}\int\Omega_{q_{i}\boldsymbol{q}}\left[  \left(
\varepsilon-\mu\right)  f-g\right]  $, respectively. For $\boldsymbol{j}%
_{neq}$, the Einstein relation is evident, which states that the electric
field and the gradient of chemical potential $\boldsymbol{\nabla}\mu/e$ are
equivalent in inducing the electric current. The Mott relation is also easy to
obtain, which reads $\boldsymbol{\alpha}^{i}=\frac{\pi^{2}}{3}\frac{k_{B}%
^{2}T}{e}\frac{\partial\boldsymbol{\sigma}^{i}\left(  \varepsilon\right)
}{\partial\varepsilon}|_{\varepsilon=\mu}$ at low temperature \cite{Xiao2006},
where $\boldsymbol{\sigma}^{i}\left(  \varepsilon\right)  $ is the Hall
conductivity at zero temperature with $\varepsilon$ the Fermi energy. As for
$\boldsymbol{j}_{eq}$, we obtain
\begin{equation}
\boldsymbol{M}=\int(f\boldsymbol{m}-e\boldsymbol{\Omega}g), \label{eq: M}%
\end{equation}
where $(\boldsymbol{\Omega})_{k}=\frac{1}{2}\Omega_{q_{i}q_{j}}\epsilon_{ijk}$
is the vector form of the antisymmetric tensor $\Omega_{\boldsymbol{q}%
\boldsymbol{q}}$, and $g=-\frac{1}{\beta}\ln[1+e^{-(\varepsilon-\mu)}]$ is the
grand potential density for a particular state. A key observation is that
$\boldsymbol{M}$ coincides with the orbital magnetization
\cite{Xiao2006,Shi2007,resta orbital magnetization}, namely the dipole density
of the electric current, hence $\boldsymbol{j}_{eq}$ is just the magnetization current.

It is important to note that, in the previous semiclassical transport approach
\cite{Xiao2006} $\boldsymbol{M}$ is obtained separately from its thermodynamic
definition $\boldsymbol{M}=-\partial G_{tot}/\partial\boldsymbol{B}$ with
$G_{tot}$ the grand potential density and $\boldsymbol{B}$ the magnetic field,
whereas the form of the magnetization current $\boldsymbol{\nabla}%
\times\boldsymbol{M}$ is known from the electrodynamics. Thereby, to
generalize this approach to other physical quantities $\boldsymbol{\theta}$,
e.g., spin and spin current, is difficult because the generalizations of the
magnetization current in these cases are not known. In fact, this is a main
theoretical difficulty in the study of the thermoelectric responses of spin
and spin current. Moreover, inhomogeneities from mechanical perturbation
fields $w$, which are usually present in practical materials, cannot be
incorporated into the previous theory \cite{Xiao2006}. On the other hand, in
the present approach both the dipole density and its contribution to the
electric current emerge automatically just through the manipulation of
$\boldsymbol{j}_{loc}$ itself, in the presence of $w$. Thus, if one can
generalize this approach to the thermoelectric responses of physical
quantities other than the electric current, the generalization of the
magnetization current can be obtained.

Applying the above new approach in these cases is not straightforward because
the perturbed wave-packet (by gradients of $w$ and $\phi$) is needed to
calculate the $\langle\Phi|\hat{\boldsymbol{\theta}}|\Phi\rangle$ term of Eq.
(\ref{eq: local density}) \cite{gao prl} (the electric current is special
since $\langle\Phi|\hat{\boldsymbol{v}}|\Phi\rangle=\dot{\boldsymbol{r}}$ is
already given by the equations of motion). Trying to overcome this difficulty,
we note that one may introduce an auxiliary coupling term $e\hat
{\boldsymbol{v}}\cdot\boldsymbol{A}$ ($\boldsymbol{A}$ is the vector
potential) in the wave-packet Lagrangian, so that $\langle\Phi|\hat
{\boldsymbol{v}}|\Phi\rangle$ and $\boldsymbol{j}_{loc}$ can also be obtained
by the variation of the action with respect to the field $\boldsymbol{A}$.
Then we are faced with a field-variational problem.

This observation stimulates the generic idea that, to obtain $\rho
_{loc}^{\boldsymbol{\theta}}\left(  \boldsymbol{r}\right)  $ we consider the
Hamiltonian:%
\begin{equation}
\hat{\mathcal{H}}=\hat{H}_{0}+\hat{\boldsymbol{\theta}}\cdot\boldsymbol{h}%
(\boldsymbol{r}_{c},t)-e\phi\left(  \boldsymbol{r}_{c}\right)  ,
\label{eq: auxillary Hamiltonian}%
\end{equation}
where $\boldsymbol{h}$ is the slowly-varying field that couples to the
considered physical observable $\hat{\boldsymbol{\theta}}$, and thus has an
unambiguous physical meaning determined by that of $\hat{\boldsymbol{\theta}}%
$. For instance, when $\hat{\boldsymbol{\theta}}$\ are the spin and electric
current, $\boldsymbol{h}$ are the Zeeman field and vector potential,
respectively. In some realizations of $\hat{\boldsymbol{\theta}}$, the
explicit form of $\boldsymbol{h}$ may not be familiar, e.g., when
$\hat{\boldsymbol{\theta}}$ is the conventional spin current $\boldsymbol{h}$
is the so-called spin-dependent vector potential \cite{Gorini2012,Dyrdal2016}.
This does not matter since knowing the explicit form of $\boldsymbol{h}$ is
not necessary in our method. This is because the auxiliary term $\hat
{\boldsymbol{\theta}}\cdot\boldsymbol{h}$ is introduced to acquire the
thermoelectric response of $\hat{\boldsymbol{\theta}}$ and is set to zero
($\boldsymbol{h}=0$) at the last of the calculation. In general, both
$\hat{\boldsymbol{\theta}}$ and $\boldsymbol{h}$ are tensors and the product
denotes the contraction between them.

Next we consider the dynamics of wave-packet $|\Phi\rangle$ constructed from
Hamiltonian $\hat{\mathcal{H}}$. The action for the wave-packet state is
\cite{Xiao2010}:%
\begin{equation}
S=\int dtL,\text{ }L=\langle\Phi|i\partial_{t}-\hat{\mathcal{H}}|\Phi\rangle,
\label{eq: action}%
\end{equation}
where $L$ is the wave-packet Lagrangian \cite{supp}. It can be easily verified
that the variation of $S$ with respect to $\langle\Phi|$ gives the Schrodinger
equation satisfied by the wave-packet. The variation with respect to
$\boldsymbol{h}$ instead gives%
\begin{equation}
\frac{\delta S}{\delta\boldsymbol{h}}|_{onshell}=-\int dt\langle\Phi
|\frac{\delta\hat{\mathcal{H}}}{\delta\boldsymbol{h}}|\Phi\rangle|_{onshell}
\label{key-action-1}%
\end{equation}
for on-shell wave-packet states (states that satisfy the Schrodinger
equation). By the definition of the field variation \cite{supp}, the right
hand side of Eq. (\ref{key-action-1})\ is simply $-\langle\Phi|\hat
{\boldsymbol{\theta}}\delta\left(  \hat{\boldsymbol{r}}-\boldsymbol{r}\right)
|\Phi\rangle$. Notice that $|\Phi\rangle$ becomes the wave-packet from the
original Hamiltonian $\hat{H}$ (\ref{eq: original Hamiltonian}) in the limit
$\boldsymbol{h}\rightarrow0$. Combining Eqs. (\ref{eq: local density original}%
) and (\ref{key-action-1}) we get the following vital relation after summing
over all wave-packets:
\begin{equation}
\rho_{loc}^{\boldsymbol{\theta}}=-\int\left[  d\mathbf{q}\right]
d\boldsymbol{r}Df\frac{\delta S}{\delta\boldsymbol{h}}|_{onshell}%
^{\boldsymbol{h}\rightarrow0}. \label{key-link-1}%
\end{equation}
In the following, we omit the label $\boldsymbol{h}\rightarrow0$ for
simplicity, but all results are evaluated in this limit.

Starting from Eq. (\ref{key-link-1}), a straightforward derivation \cite{supp}
yields the important result
\begin{equation}
\rho_{loc}^{\boldsymbol{\theta}}=\int Df\left(  \partial_{\boldsymbol{h}%
}\varepsilon_{tot}-\Omega_{\boldsymbol{h}\mathcal{T}}\right)  -\partial
_{r^{i}}\int fm^{i\boldsymbol{\theta}}. \label{key-field}%
\end{equation}
We note that this equation indicates $\langle\Phi|\hat{\boldsymbol{\theta}%
}|\Phi\rangle=\partial_{\boldsymbol{h}}\varepsilon_{tot}-\Omega
_{\boldsymbol{h}\mathcal{T}}$ with $\Omega_{\boldsymbol{h}\mathcal{T}}%
=\Omega_{\boldsymbol{h}\boldsymbol{r}}\dot{\boldsymbol{r}}+\Omega
_{\boldsymbol{h}\boldsymbol{q}}\dot{\boldsymbol{q}}+\Omega_{\boldsymbol{h}t}$.
Notwithstanding the similar form to Eq. (\ref{eq: local j_e}), there is a
basic difference: the $\boldsymbol{q}$ derivative in Eq. (\ref{eq: local j_e})
is replaced by the derivative with respect to the field $\boldsymbol{h}$ that
couples to the considered observable $\boldsymbol{\theta}$. In fact, Eq.
(\ref{eq: local j_e}) can be reinterpreted from the view point of the
field-variation as the special case of Eq. (\ref{key-field}) when
$\hat{\boldsymbol{\theta}}=-e\hat{\boldsymbol{v}}$ and $\boldsymbol{h}%
=-\boldsymbol{A}$: Since the vector potential is always minimally coupled into
the Hamiltonian in the combined form $\boldsymbol{q}+e\boldsymbol{A}$, the
$\boldsymbol{h}$ derivative is proportional to the $\boldsymbol{q}$ derivative
with a factor $-e$.

The dipole moment of $\boldsymbol{\theta}$ takes the form of
$m^{i\boldsymbol{\theta}}=\operatorname{Im}\langle\partial_{q_{i}%
}u|\varepsilon-\hat{\mathcal{H}}|\partial_{\boldsymbol{h}}u\rangle$. It is
related to the gradient correction of the wave-packet energy in the way that
$\Delta\varepsilon=\operatorname{Im}\langle\partial_{q_{i}}u|\varepsilon
-\hat{\mathcal{H}}|\partial_{w}u\rangle\partial_{r^{i}}%
w+m^{i\boldsymbol{\theta}}\cdot\partial_{r^{i}}\boldsymbol{h}$. Thus the
gradient correction can be generally interpreted as the potential energy of
the dipole moment in an external field.

Starting from Eq. (\ref{key-field}) and taking some technical steps similar to
those from Eq. (\ref{eq: local j_e}) to Eq. (\ref{j}) \cite{supp}, we obtain%
\begin{align}
\rho_{eq}^{\boldsymbol{\theta}}=  &  \partial_{\boldsymbol{h}}G_{tot}%
-\partial_{i}M^{i\boldsymbol{\theta}},\\
\rho_{neq}^{\boldsymbol{\theta}}=  &  \sigma^{i\boldsymbol{\theta}}\left(
E_{i}+\partial_{i}\mu/e\right)  -\alpha^{i\boldsymbol{\theta}}\partial_{i}T.
\label{eq: eq and neq general}%
\end{align}
Here $\rho_{eq}^{\boldsymbol{\theta}}$ is the equilibrium part. In the case of
the electric current, its first term vanishes since the $\boldsymbol{q}$
variable has already been integrated out, and its second term gives the
magnetization current. $G_{tot}=\int Dg\left(  \varepsilon_{tot}\right)
=G+\Delta G$, where $G=\int g(\varepsilon)$ is the local part and $\Delta
G=\int\left[  f\Delta\varepsilon+\Omega_{q_{i}r_{i}}g\right]  $ is induced by
inhomogeneity.%
\begin{equation}
M^{i\boldsymbol{\theta}}=\int\left(  fm^{i\boldsymbol{\theta}}+g\Omega
_{q_{i}\boldsymbol{h}}\right)  \label{M general}%
\end{equation}
is recognized as the dipole density of $\boldsymbol{\theta}$ since%
\begin{equation}
M^{i\boldsymbol{\theta}}=\frac{\partial G_{tot}}{\partial\left(
\partial_{r_{i}}\boldsymbol{h}\right)  },
\end{equation}
which is the thermodynamical definition of the dipole density of a physical
quantity. This definition reduces to the orbital magnetization
\cite{Shi2007,Xiao2006} and the spin dipole density (whose antisymmetric part
is called spin toroidization) \cite{Gao2018,Shitade2018} when
$\boldsymbol{\theta}$ is the electric current ($\boldsymbol{h}=-\boldsymbol{A}%
$) and the spin ($\boldsymbol{h}$ is the Zeeman field), respectively. The fact
that the divergence of $M^{i\boldsymbol{\theta}}$ contributes to the
$\boldsymbol{\theta}$ density also verifies its physical meaning.

Equation (\ref{eq: eq and neq general}) describes the general linear response
to the electric field and statistical force, with the coefficients%
\begin{equation}
\sigma^{i\boldsymbol{\theta}}=-e\int f\Omega_{q_{i}\boldsymbol{h}},\text{
}\alpha^{i\boldsymbol{\theta}}=\frac{1}{T}\int\Omega_{q_{i}\boldsymbol{h}%
}[(\varepsilon-\mu)f(\varepsilon)-g(\varepsilon)].
\end{equation}
The Einstein relation is apparent in Eq. (\ref{eq: eq and neq general}). The
generalized Mott relation can be also proved \cite{supp}:
\begin{equation}
\alpha^{i\boldsymbol{\theta}}=\frac{1}{e}\int d\varepsilon\frac{\partial
f}{\partial\varepsilon}\frac{\varepsilon-\mu}{T}\sigma^{i\boldsymbol{\theta}%
}\left(  \varepsilon\right)  , \label{Mott}%
\end{equation}
where $\sigma^{i\boldsymbol{\theta}}\left(  \varepsilon\right)  $ is the
zero-temperature value of $\sigma^{i\boldsymbol{\theta}}$ with Fermi energy
$\varepsilon$. At low temperatures much less than the distances between the
chemical potential and band edges, the Sommerfeld expansion is legitimate
\cite{Xiao2016PRB}, yielding the standard Mott relation $\alpha
^{i\boldsymbol{\theta}}=-\frac{\pi^{2}k_{B}^{2}T}{3e}\frac{\partial
\sigma^{i\boldsymbol{\theta}}\left(  \varepsilon\right)  }{\partial
\varepsilon}|_{\varepsilon=\mu}$, which relates $\alpha^{i\boldsymbol{\theta}%
}$ to the energy derivative of $\sigma^{i\boldsymbol{\theta}}$ around the
chemical potential.

For the convenience of calculation, one can express the dipole moment and
Berry curvatures involving $\boldsymbol{h}$ derivatives in a more explicit
form:
\begin{align}
m^{i\boldsymbol{\theta}}  &  =\operatorname{Im}\sum_{m\neq n}\frac{\langle
u_{n}|\hat{v}^{i}|u_{m}\rangle\langle u_{m}|\hat{\boldsymbol{\theta}}%
|u_{n}\rangle}{\varepsilon_{n}-\varepsilon_{m}},\nonumber\\
\Omega_{q_{i}\boldsymbol{h}}  &  =-2\operatorname{Im}\sum_{m\neq n}%
\frac{\langle u_{n}|\hat{v}^{i}|u_{m}\rangle\langle u_{m}|\hat
{\boldsymbol{\theta}}|u_{n}\rangle}{\left(  \varepsilon_{n}-\varepsilon
_{m}\right)  ^{2}}, \label{ab initio}%
\end{align}
where $n$ is the index of the band we are considering. In obtaining these two
expressions the $\boldsymbol{h}$ derivatives have been done, followed by
setting $\boldsymbol{h}=0$, thus both terms exist only if $\hat
{\boldsymbol{\theta}}$ does not commute with the genuine Hamiltonian $\hat
{H}_{0}$. Therefore, the dipole density and the linear response coefficients
we discussed before is a property pertaining to such \textquotedblleft
nonconserved\textquotedblright\ quantities. It is also worthwhile to mention
that our results apply to any operator $\hat{\boldsymbol{\theta}}$ that is
well-defined in the Bloch representation. For the conventional spin current
operator, $\Omega_{q_{i}\boldsymbol{h}}$ is just the quantity sometimes
referred to as the \textquotedblleft spin Berry curvature\textquotedblright%
\ in first-principles literatures \cite{Yan2016,yao zhong fang}.

Having identified the generalization of the magnetization current
$-\partial_{i}M^{i\boldsymbol{\theta}}$, we can now understand the
thermoelectric response of $\boldsymbol{\theta}$ in a direct way when
inhomogeneities come only from temperature and chemical potential
\cite{Xiao2006}. In this simple case the local density
(\ref{eq: local density}) reduces to
\begin{equation}
\rho_{loc}^{\boldsymbol{\theta}}=\int f\langle u|\hat{\boldsymbol{\theta}%
}|u\rangle+\sigma^{i\boldsymbol{\theta}}E_{i}-\partial_{r^{i}}\int
fm^{i\boldsymbol{\theta}},
\end{equation}
where the $\sigma^{i\boldsymbol{\theta}}E_{i}$ term arises from the interband
mixing of Bloch states induced by the electric field
\cite{Culcer2004,Xiao2017}, whereas
\begin{equation}
-\partial_{i}M^{i\boldsymbol{\theta}}=-\partial_{r^{i}}\int
fm^{i\boldsymbol{\theta}}-\sigma^{i\boldsymbol{\theta}}\partial_{i}%
\mu/e+\alpha^{i\boldsymbol{\theta}}\partial_{i}T.
\end{equation}
Hence the nonequilibrium part of $\rho_{loc}^{\boldsymbol{\theta}}-\left(
-\partial_{i}M^{i\boldsymbol{\theta}}\right)  $, which corresponds to the
subtraction of the magnetization current from the local electric
current\ density in \cite{Xiao2006}, just gives the thermoelectric response
satisfying the Einstein and Mott relations. In this picture, contributions
from the dipole moment $m^{i\boldsymbol{\theta}}$ cancel out in the linear
response, while the Berry-phase correction to the dipole density plays the
vital role in validating both relations.

$\sigma^{i\boldsymbol{\theta}}$ contains a Streda term $\sigma
^{i\boldsymbol{\theta},\text{II}}$ \cite{Sinova2015,Nagaosa2010,Ebert2015}
whose zero temperature value is related to the dipole density as:
\begin{equation}
M^{i\boldsymbol{\theta}}=\frac{1}{e}\int d\varepsilon f\left(  \varepsilon
\right)  \sigma^{i\boldsymbol{\theta},\text{II}}\left(  \varepsilon\right)  .
\label{M Streda}%
\end{equation}
This relation can be derived from Eq. (\ref{M general}) by the same procedure
in \cite{note M-Streda}. $\sigma^{i\boldsymbol{\theta},\text{II}}\left(
\varepsilon\right)  $ has the following form \cite{Ebert2015}:
\begin{equation}
\sigma^{i\boldsymbol{\theta},\text{II}}(\varepsilon)=\operatorname{Re}%
\int_{-\infty}^{\varepsilon}\frac{d\epsilon}{2\pi}\text{Tr}[\hat
{\boldsymbol{\theta}}\hat{G}^{R}\hat{\jmath}_{e}^{i}\frac{d\hat{G}^{R}%
}{d\epsilon}-\hat{\boldsymbol{\theta}}\frac{d\hat{G}^{R}}{d\epsilon}%
\hat{\jmath}_{e}^{i}\hat{G}^{R}].
\end{equation}
Here $\hat{G}^{R}$ is the bare retarded Green's function. This connection is
useful in model calculations. For instance, in the two-dimensional Rashba
model with both Rashba subbands partially occupied \cite{Sinova2015}, the
zero-temperature Streda term of the conventional spin Hall conductivity is
$\sigma_{\boldsymbol{s}}^{xy,\text{II}}\left(  \varepsilon\right)  =\frac
{-e}{8\pi}\left[  \frac{k_{R}}{k_{0}\left(  \varepsilon\right)  }-\frac
{k_{0}\left(  \varepsilon\right)  }{k_{R}}\right]  \Theta\left(
-\varepsilon\right)  $, where $\Theta$ is the step function, $k_{0}\left(
\varepsilon\right)  =\alpha_{R}^{-1}\sqrt{\varepsilon_{R}^{2}+2\varepsilon
_{R}\varepsilon}$ with $\alpha_{R}$ the Rashba coefficient, $k_{R}=m\alpha
_{R}/\hbar^{2}$ ($m$ is the effective mass) the Rashba wave-vector and
$\varepsilon_{R}=\alpha_{R}k_{R}$ the Rashba energy. Thus the zero-temperature
dipole density of the conventional spin current is obtained as
$M_{\boldsymbol{s}}^{xy}=-\frac{\epsilon_{R}}{12\pi}$.

Finally, for completeness, we demonstrate that the Einstein and Mott relations
still hold in the presence of elastic scattering on weak static disorder. As
mentioned before, the total local density has a term $\delta\rho
_{loc}^{\boldsymbol{\theta}}=\int\delta f\langle\Phi|\hat{\boldsymbol{\theta}%
}|\Phi\rangle$. $\delta f$ in steady states is determined by the linearized
Boltzmann equation \cite{Ziman1960} ($P_{\boldsymbol{kq}}$ is the scattering
rate in the Born approximation)
\begin{equation}
\dot{\boldsymbol{r}}\cdot\partial_{\boldsymbol{r}}f+\dot{\boldsymbol{k}}%
\cdot\partial_{\boldsymbol{k}}f=\int P_{\boldsymbol{kq}}\left(  f_{tot}%
(\boldsymbol{q})-f_{tot}(\boldsymbol{k})\right)
.\label{eq: Boltzman equation}%
\end{equation}
The left hand side is simply $\boldsymbol{F}\cdot\partial_{\boldsymbol{k}}f$,
where $\boldsymbol{F}=-e\boldsymbol{E}-\mathbf{\partial}_{\boldsymbol{r}}%
\mu-\frac{\varepsilon_{0}-\mu}{T}\mathbf{\partial}_{\boldsymbol{r}}T$
\cite{Ziman1960}. Thus $\delta f\propto\boldsymbol{F}\cdot\partial
_{\boldsymbol{k}}f$ in the linear response \cite{deltaf}, validating the
Einstein and Mott relations \cite{Ashcroft}. In systems with Berry-phase
corrections, it is well known that two extrinsic effects called skew
scattering and coordinate-shift need also be incorporated into the Boltzmann
equation \cite{Sinitsyn2008,Akera2013}. We show in Supplemental Material \cite{supp}
that these two effects do not break the Einstein and Mott relations. Besides
modifying the occupation function, disorder also alters $\langle\Phi
|\hat{\boldsymbol{\theta}}|\Phi\rangle$ by inducing interband mixing of Bloch
states \cite{Xiao2017}. This contribution, known as side-jump velocity for
$\hat{\boldsymbol{\theta}}=\hat{\boldsymbol{v}}$ \cite{Xiao2017,Sinitsyn2006},
is averaged by $\delta f\propto\boldsymbol{F}\cdot\partial_{\boldsymbol{k}}f$,
hence does not go against the Einstein or Mott relation.

The proposed approach provides a unified description for the
anomalous and spin Nernst effects, the thermally induced spin and
spin-orbit torque \cite{Freimuth2014}. It also applies to or can be
extended in several directions of current great interest. First, it can be
further generalized to a framework of linear thermoelectric responses of
dipole densities. In the present paper we are
limited to linear responses of operator $\hat{\boldsymbol{\theta}}$ that is
well-defined in periodic crystals, and only the equilibrium dipole density of
$\hat{\boldsymbol{\theta}}$ is needed. We extend \cite{note} the variational approach
to nonlinear responses of $\hat{\boldsymbol{\theta}}$ and
then obtain the linear response of $\hat{\boldsymbol{\theta}}$-dipole.
This extension enables to, for example, analyze the temperature gradient induced
orbital magnetization, thus paving the way for thermal generation and control
of magnetization via the orbital degree of freedom, which is especially
important in low-symmetry valley systems. Second, by
allowing for the second order spatial derivative of the $\boldsymbol{h}$ field,
the variational approach yields a general theory for various quadrupole densities
($r_{i}r_{j}\theta_{l}$), such as the orbital magnetic quadrupole
\cite{Shitade2018-2} which serves as a order parameter
of systems with combined time-reversal and inversion symmetry. Third, the
generalization into the case of degenerate bands, i..e, a non-abelian
formalism \cite{Cheng2012}, can also be pursued. Fourth, the field-variational
approach applies to bosonic systems as well. Indeed the idea of our work has
been shown recently to work in the thermal spin generation and spin
Nernst effect of magnons in noncollinear antiferromagnetic insulators
\cite{Li2019}.

\begin{acknowledgments}
We thank F. Freimuth for useful discussions.
Q.N. is supported by DOE (DE-FG03-02ER45958, Division of Materials
Science and Engineering) on the geometric formulation in this work.
L.D., C.X. and B.X. are supported by NSF (EFMA-1641101) and Welch Foundation
(F-1255).
\end{acknowledgments}


\begin{thebibliography}{99}
\bibitem {note}Since this extension contains rich physics, it will be
discussed in separate works.
\end{thebibliography}
\end{document}